\RequirePackage{ifpdf}
\ifpdf % We are running pdfTeX in pdf mode
\documentclass[pdftex]{sigma}
\else
\documentclass{sigma}
\fi

\numberwithin{equation}{section}

\begin{document}

\allowdisplaybreaks

\renewcommand{\thefootnote}{$\star$}

\renewcommand{\PaperNumber}{100}

\FirstPageHeading

\ShortArticleName{Families of Integrable Equations}

\ArticleName{Families of Integrable Equations\footnote{This
paper is a contribution to the Proceedings of the Conference ``Symmetries and Integrability of Dif\/ference Equations (SIDE-9)'' (June 14--18, 2010, Varna, Bulgaria). The full collection is available at \href{http://www.emis.de/journals/SIGMA/SIDE-9.html}{http://www.emis.de/journals/SIGMA/SIDE-9.html}}}

\Author{Pavlos KASSOTAKIS~$^\dag$ and Maciej NIESZPORSKI~$^\ddag$}

\AuthorNameForHeading{P.~Kassotakis and M.~Nieszporski}

\Address{$^\dag$~Department of Mathematics and Statistics University of Cyprus,\\
\hphantom{$^\dag$}~P.O.~Box: 20537, 1678 Nicosia, Cyprus}
\EmailD{\href{mailto:kassotakis.pavlos@ucy.ac.cy}{kassotakis.pavlos@ucy.ac.cy}, \href{mailto:pavlos1978@gmail.com}{pavlos1978@gmail.com}}

\Address{$^\ddag$~Katedra Metod Matematycznych Fizyki, Uniwersytet Warszawski,\\
\hphantom{$^\ddag$}~ul. Ho\.za 74, 00-682 Warszawa, Poland}
\EmailD{\href{mailto:maciejun@fuw.edu.pl}{maciejun@fuw.edu.pl}}

\ArticleDates{Received May 23, 2011, in f\/inal form October 20, 2011;  Published online October 28, 2011}

\Abstract{We present a method to obtain families of lattice equations. Specif\/ically we focus on two of such families, which include 3-parameters and their members are connected through B\"acklund transformations. At least one of the members of each family is integrable, hence the whole family inherits some integrability properties.}

\Keywords{integrable lattice equations; Yang--Baxter maps; consistency around the cube}

\Classification{82B20; 37K35; 39A05}

\section{Introduction}
%%%%%%%%%%%%%%%%%%%%%%

 Discrete mathematics returned on the interest of mathematicians at
 the beginning of the $20^{th}$ century. Poincar\'e, Birkhof\/f, Ritt (1924) \cite{ritt}, Julia, Fatou (1918--1923)
 \cite{julia,fatou} and many others   saw the necessity of exploring the discrete  scene.
 Unfortunately, this trend was paused through the two big wars and  only after 1960, keeping pace with the revolution
 caused  by the discovery of soliton from Zabusky and Kruskal~\cite{kru1}, mathematicians started to investigate  discrete systems in the context of integrable systems.

It was the work of Hirota \cite{hirota-0}, as well as Ablowitz et al.~\cite{ablo-ladik} and separately Capel and his school~\cite{nij-qui-cap}, which introduced
lattice and dif\/ferential dif\/ference analogues of many integrable PDE's. The introduction of discrete versions of integrable ODE's, surprisingly, came later with the QRT family of mappings by Quispel, Roberts and Thomson~\cite{qrt1} and by the work of  Papageorgiou et al.~\cite{pap1,nij55},  where  Liouville  integrable maps~\cite{ves2} were obtained by imposing   periodic staircase initial data on  integrable lattices. Another way to obtain integrable mappings from an integrable lattice equation  was suggested in series of papers~\cite{ABSf,pap2,BoSu-book}.  Actually with this procedure one can get involutive mappings (composition of the map with itself is the identity map)  which are set theoretical solutions of  the quantum Yang--Baxter equation the so called Yang--Baxter maps~\cite{drin,Ves,ABSf,PSTV}.
As in our previous work~\cite{KaNie}, we focus here on the inverse  procedure, i.e.\ how to  obtain integrable lattice equations from involutive mappings that may or may not satisfy the Yang--Baxter equation.

The main result of the paper is that the procedure we have in mind can lead to families of equations.
It is necessary to mention that points of the lattice
may not be related in a unique way.
The members of families are related by a B\"acklund transformation (see Section~\ref{bt}) and since in considered cases at least one of the members is integrable, the whole family inherits some properties from the distinguished member.
 We stress that solutions of each member of the family can be obtained from solutions of the integrable member by  discrete quadratures
(which can be regarded as sort of B\"acklund transformation) and
in this sense each member of the family is integrable. However, we discuss here  hallmarks of integrability of the members of the family such as consistency around the cube property or $\tau$-function formulation.
Notion of the family of discrete integrable systems should not be confused with notion of hierarchies of integrable systems. The later notion was widely investigated in the literature whereas for the former one we can indicate only the articles that investigate the family of discrete KdV equations~\cite{NijOht} and the family of discrete Boussinesq equations~\cite{Ni1,Ni2,Ni0}.

We discuss here two examples, the f\/irst one is continuation of our previous paper~\cite{KaNie}. We introduce a family of dif\/ference equations associated
with type III of maps discussed in \cite{ABSf,pap2} (we introduced families related to types IV and V in~\cite{KaNie}).
Example of the map of type III is a~map $\mathbb{C}^2\ni (u,v) \mapsto (U,V) \in \mathbb{C}^2$
\begin{gather*}
%\label{UV}
U =  v \frac{p u - q v}{q u - p v} , \qquad
V = u \frac{p u - q v}{q u - p v}.
\end{gather*}
and the three parameter family of  equations (see Section~\ref{out}) reads
\begin{gather}
\label{ip}
\psi_{12}=\psi+{a \ln \frac{pu-qv}{qu-pv} +\left(p^2-q^2\right) \left[ b \frac{uv}{qu-pv} -c \frac{1}{pu-qv} \right]},
\end{gather}
where $u$ and $v$ are given implicitly by
\begin{gather}
\label{ipd}
a \ln u+  p\left(b u+c \frac{1}{u}\right)=\psi _1+\psi , \qquad
a \ln v+  q \left(b v+c \frac{1}{v}\right)=\psi _2+\psi,
\end{gather}
function $\psi$ is dependent variable on ${\mathbb Z}^2$ and we denote $\psi(m,n)=:\psi$,  $\psi(m+1,n)=:\psi_1$, $\psi(m,n+1)=:\psi_2$,  $\psi(m+1,n+1)=:\psi_{12}$, $p:=p(m)$ and $q:=q(n)$ are given functions of a~single variable
and $a$, $b$ and $c$  are arbitrary constants (we assume that one of the constants~$a$,~$b$ or~$c$ is not equal to zero).
However, ought to possible branching in formulas \eqref{ipd}, the system~\eqref{ip},~\eqref{ipd}  needs  specifying (as it was pointed us by Professors Frank Nijhof\/f and Yuri Suris). The specif\/ication is achieved by demanding that functions $u$ and $v$ obey
\begin{gather}
\label{u2v1}
u_2 =  v \frac{p u - q v}{q u - p v} , \qquad
v_1 = u \frac{p u - q v}{q u - p v}.
\end{gather}
After this specif\/ication there is still some freedom left in f\/inding solutions of~\eqref{ip}, \eqref{ipd} for  given initial conditions on $\psi$. The freedom lies in f\/inding the initial conditions for~$u$ and~$v$ out of initial conditions on $\psi$  by means of~\eqref{ipd}. The solution need not to be unique.

All the equations within the family are consistent around the cube (for the  consistency around the cube property see \cite{FN,FN-KN,ABS,Boll}, notice we resign from multiaf\/f\/inity assumption of paper~\cite{ABS}).
 We f\/ind especially interesting the fact that we obtain examples of lattice equations together with transformations which can be regarded as B\"acklund transformations but not in the usual sense; we usually require B\"acklund transformation to be linearisable  (see Def\/inition~\ref{BTdef}) and this requirement is violated in these examples.
Therefore Lax pair could not be easily found from this sort of B\"acklund transformation
and it is not clear if the Lax pair  exists in these cases.
Members of the family are  Hirota's sine-Gordon equation (choice of parameters $b=0=c$) referred also to as lattice potential modif\/ied KdV \cite{Hir-sG,NC,FaVo,BoPi,Zab,BoSu-book,RG2-Sol}
(see Section~\ref{tran} where we discuss  various forms of lattice equations)
\begin{gather*}
%\label{ix}
 p(xx_1 + x_2x_{12})=q(xx_2 + x_1x_{12})
\end{gather*}
and lattice Schwarzian KdV \cite{NC}  in a disguise, see Section~\ref{tran} (choice of parameters $a=0=b$ or $a=0=c$)
\begin{gather*}
%\label{iy}
 p^2(y_{12} + y_1)(y_2 + y)=q^2(y_{12}+y_2)(y_1+y).
\end{gather*}

In the second example we go away from the maps of papers \cite{ABSf,pap2} and consider the map
\begin{gather*} %\label{iH}
U=v+k\left(1-\frac{v}{u}\right), \qquad V=u+k\left(-1+\frac{u}{v}\right),
\end{gather*}
which gives also a 3 parameter family of equations (see Section~\ref{hero}) including
Hirota's KdV lattice equation~\cite{hirota-0}
\begin{gather*} %\label{iK}
x_{12}-x=\kappa\left(\frac{1}{x_2}-\frac{1}{x_1}\right)
\end{gather*}
and two further bilinear equations
\begin{gather*}
y_1y-y_{12}y_1=\kappa(y_{12}y+y_1y_2),        \qquad
z_{12}z+z_1z_2=z_{12}z_2+z_{12}z_1.
 \end{gather*}
In this case an interesting fact is that the procedure  yields $\tau$-function representation of the family (see e.g.~\cite{NijOht})
\begin{gather*}
\tau_{112}\tau -\kappa \tau_{11}\tau_{2}=\tau_{12}\tau_{1},\qquad
\tau_{122}\tau +\kappa \tau_{22}\tau_{1}=\tau_{12}\tau_{2}.
\end{gather*}
In Section~\ref{tran}, we give an overview of point transformations, B\"acklund transformations and dif\/ference substitutions and touch the issue of equivalence of lattice equations. We proceed in Section~\ref{out} where we present the method that leads to families of lattice equations.
In Section~\ref{maps} we relate our f\/indings to some results of the papers~\cite{ABSf,pap2}, followed by Section~\ref{hero} where  we deal with Hirota's KdV lattice equation.
Then we explain how to get B\"acklund transformation between members of the families (Section~\ref{bt}) and we end the paper with some conclusions and perspectives for future work.

\section{Point transformations, dif\/ference substitutions,  B\"acklund\\ transformations and equivalence of lattice equations}
\label{tran}

Before we start we would like to give some  def\/initions and recall some well known relations~\cite{FaVo,BoPi,Zab,NijOht,RG2-Sol} between equations that appear in the article (terminology used by various authors is far from being unif\/ied). Let us consider~$k$ dependent variables of $n$ independent ones: $u^i(m^1,\ldots,m^n)$, $i=1,\ldots,k$. We  denote $M\equiv (m^1,\ldots,m^n)$.

\begin{definition}[change of independent variables]
{\em By change of independent variables} we understand the bijection $f: \mathbb{Z}^n \rightarrow \mathbb{Z}^n$
\[\tilde{m}^i=f^i(M), \qquad i=1,\ldots,n. \]
\end{definition}

 2D examples are $\tilde{m}^1=m^1$, $\tilde{m}^2=m^1+m^2$, or $\tilde{m}^1=m^1+2 m^2$, $\tilde{m}^2=m^1+m^2.$
\begin{definition}[point transformations not altering independent variables]
 By {\em point transformation not altering independent variables} we understand an invertible map $F$ between subsets of~$\mathbb{C}^k$
\[
\tilde{u}^i(M) =F^i\big(u^1(M) ,\ldots, u^k(M) ; M\big), \qquad i=1,\ldots,k.
\]
\end{definition}

\begin{definition}[equivalence of lattice equations]
Two lattice  equations are {\em equivalent} if and only if there exists composition of  point transformation with change of independent variables which maps solutions of one equation to solutions  of the second one.
\end{definition}
Examples of various disguises of the same equation are
\begin{itemize}\itemsep=0pt
\item Hirota's sine-Gordon  equation
\[q \sin (\psi_{12} + \psi -\psi_{1} -\psi_{2})=
p \sin (\psi_{12} + \psi +\psi_{1} +\psi_{2})\]
turns  into
\begin{gather}\label{H3}
 (H3^0): \quad p (xx_1 + x_2x_{12}) = q (xx_2 + x_1x_{12})
\end{gather}
$H3^0$ equation from ABS list  \cite{ABS} by means of point transformation $x=i^{m+n}e^{2i(-1)^n \psi}$.
 $H3^0$ in  turn  can be transformed into  lattice potential modif\/ied KdV
\[
 p (ww_1 - w_2w_{12}) = q (ww_2 - w_1w_{12})
\]
by  substitution $x=i^{m+n} w$.
\item Schwarzian KdV equation (or cross ratio equation, or equation $Q1^0$ on ABS list)
\[\frac{(z_{12}-z_1)(z_2-z)}{(z_{12}-z_2)(z_1-z)}=\frac{q^2}{p^2}\]
under the point transformation $z=(-1)^{m+n} y$ turns into
\begin{gather}
\label{A1}
(A1^0): \quad p^2(y_{12} + y_1)(y_2 + y)=q^2(y_{12}+y_2)(y_1+y)
\end{gather}
which in the paper~\cite{ABS} got its own name $A1^0$.
\end{itemize}

\begin{definition}[dif\/ference substitutions]
Let $j$ points $M^i$, $ i=1,\ldots,j$ of a lattice are given. By {\em difference substitution of order $j$}  we understand a transformation
\[
\tilde{u}^i(M) =F^i\big(u^1\big(M^1\big) ,\ldots, u^k\big(M^1\big),\ldots,u^1\big(M^j\big) ,\ldots, u^k\big(M^j\big) ; M\big), \qquad i=1,\ldots,k.
\]
\end{definition}
Every point transformation  is dif\/ference substitutions of order 1. Standard examples of dif\/ference substitution (of order~2,~3 and~4 respectively)
are
\begin{itemize}\itemsep=0pt
\item
potential relation
\[v= \frac{1}{\alpha-\beta}(u_2-u_1)\]
between  lattice potential KdV
\[(u_{12}-u)(u_{1}-u_{2})=\alpha^2-\beta^2\]
and Hirota's dif\/ference KdV
\[
v_{12}-v=\frac{\alpha+\beta}{\alpha-\beta} \left(\frac{1}{v_{1}}-\frac{1}{v_{2}}\right);
\]
\item
Miura-type transformation
\[v= \frac{\beta \psi _2 - \alpha \psi _1}{(\beta  - \alpha) \psi } \]
between  $H3^0$ (Hirota's sine-Gordon or lattice modif\/ied potential KdV)
\[
 \alpha (\psi_2 \psi_{12}  - \psi \psi_1 )=\beta (\psi_1 \psi_{12}-\psi \psi_2)
\]
and Hirota's dif\/ference KdV;
\item and f\/inally the introduction of $\tau$ function
\[
v=\frac{\tau_{12} \tau }{\tau_{1}\tau_{2}},
\]
which transform every solution of the compatible system
\[
\tau_{112}\tau -\kappa \tau_{11}\tau_{2}=\tau_{12}\tau_{1}, \qquad
\tau_{122}\tau +\kappa \tau_{22}\tau_{1}=\tau_{12}\tau_{2}
\] to solution of Hirota's dif\/ference KdV.
\end{itemize}
To the end we propose draft def\/inition of  B\"acklund transformation which is convenient for our purposes. However we are aware that the def\/inition is not exhaustive (some transformation that deserve this name can be not covered by the def\/inition).
\begin{definition}[B\"acklund transformations (in narrow sense)]\sloppy
\label{BTdef}
By {\em B\"acklund transformation} we understand here a transformation  between two equations $F(u_{12},u_1,u_2,u)=0$ and
$ \tilde{F}( \tilde{u}_{12}, \tilde{u}_1, \tilde{u}_2, \tilde{u})=0$
\[ \tilde{u}_1=f(\tilde{u},u, u_1), \qquad  \tilde{u}_2=g(\tilde{u},u, u_2),
\]
which is  invertible to
\[ u_1= \tilde{f}(u,\tilde{u}, \tilde{u}_1), \qquad  u_2=\tilde{g}(u,\tilde{u},\tilde{u}_2), \]
  where functions $f$ and $g$ are fractional linear in $\tilde{u}$ and functions $\tilde{f},$ $\tilde{g}$ are function fractional linear in $u$.
\end{definition}

A classical  example of B\"acklund transformation
between
\[ p(xx_1 + x_2x_{12})-q(xx_2 + x_1x_{12}) = 0\]
and
\[ p^2(y_{12}+y_1)(y_2+y)=q^2(y_{12}+y_2)(y_1+y)\]
is the transformation
\begin{gather}
\label{xy}
 y_1+y=p x_1 x,  \qquad y_2+y=q x_2 x.
\end{gather}

\section{Outline of the method}
\label{out}

We consider the $\mathbb{Z}^2$ lattice together with its horizontal edges (which can be viewed as set of ordered pair of points  of
$\mathbb{Z}^2$, i.e.\ $E_h=\left\{ ((m,n),(m+1,n))| (m,n) \in \mathbb{Z}^2 \right\}$)
and the vertical ones ($E_v=\left\{ ((m,n),(m,n+1))| (m,n) \in \mathbb{Z}^2 \right\}$).
We take into account a function $u$ which is given on horizontal edges $u: E_h \rightarrow \mathbb{C}$ and a function $v$ given on vertical ones $v: E_v \rightarrow \mathbb{C}$. Shift opera\-tors~$T_1$ and~$T_2$ act on horizontal edges in standard way $T_1 ((m,n),(m+1,n)):= ((m+1,n),(m+2,n))$,  $T_2 ((m,n),(m+1,n)):= ((m,n+1),(m+1,n+1))$ (and similarly for vertical edges). We use convention to denote shift action on a function by subscripts $T_1 u:=u_1$.

Now, the outline of the method we
developed in \cite{KaNie} can be presented as follows.

\subsection{From equations to involutive maps. Idea system}

Take a function $x$ given on vertices of the lattice and which obeys $H3^0$ equation
\begin{gather}
\label{x}
 p(xx_1 + x_2x_{12})=q(xx_2 + x_1x_{12}).
\end{gather}
Introduce f\/ields $u$ and $v$ given on horizontal and vertical edges respectively
(f\/ields u and v are actually the invariants of a symmetry group of the lattice equation (\ref{x}) as it was shown in~\cite{pap2})
\begin{gather*}
u=x x_1, \qquad v =x x_2.
\end{gather*}
We get
\begin{gather*}
 u_2u = v_1 v, \qquad  p(u_2+u)=q(v_1 +v )
\end{gather*}
and we arrive at the system of equations
\begin{gather}
\label{uv}
u_2 =  v \frac{p u - q v}{q u - p v} , \qquad
v_1 =  u \frac{p u - q v}{q u - p v}.
\end{gather}
The main idea is  to investigate system \eqref{uv} rather than equation \eqref{x} itself.
We dare to refer to the system \eqref{uv} as to 2D Idea system III.
The point is that the system \eqref{uv} admits, as we shall see, three parameter family of potentials $\psi$ given on vertices of the lattice.
Every ``potential image'' of  \eqref{uv} we refer to as idolon (adopting Plato terminology of Ideas and idolons).

\begin{figure}[th!]
\centering
\includegraphics{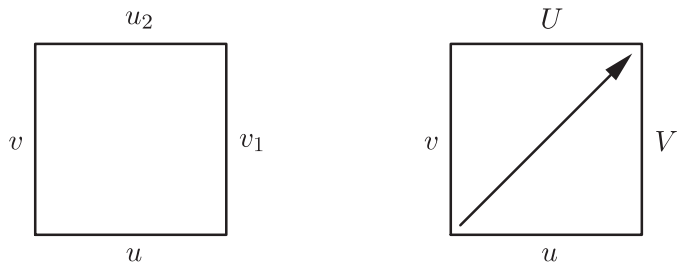}
\caption{Variables on edges of a $\mathbb{Z}^2$ lattice (left picture) and  arguments and values of a $\mathbb{C}^2 \mapsto \mathbb{C}^2 $ map (right picture).}\label{fig:c2-c2}
\end{figure}

First we apply
the standard procedure for reinterpretation of  equations on a lattice  as a map. The reinterpretation is based on identif\/ication (see Fig.~\ref{fig:c2-c2})
\begin{gather}
\label{id}
u(m,n)=u,\qquad v(m,n)=v, \qquad u(m,n+1)=U, \qquad v(m+1,n)=V,
\end{gather}
which turns system \eqref{uv} into $\mathbb{C}^2 \rightarrow \mathbb{C}^2$ map
\begin{gather}
\label{F3}
U = v \frac{p u - q v}{q u - p v}, \qquad
V = u \frac{p u - q v}{q u - p v}.
\end{gather}

We arrive at an involutive Yang--Baxter map that belongs to family of maps denoted by $F_{\rm III}$ (see \cite{ABSf}).

\subsection[Finding functions such that $F(U)+G(V)=f(u)+g(v)$]{Finding functions such that $\boldsymbol{F(U)+G(V)=f(u)+g(v)}$}

The next step is to f\/ind such functions $F$ and $G$ such that for the map \eqref{F3}
\begin{gather}
\label{FG}
F(U)+G(V)=f(u)+g(v).
\end{gather}
holds. Anticipating facts, the functions will allow us to introduce a family of potentials in the next subsection.
Dif\/ferentiation of \eqref{FG} with respect to $u$ and $v$ yields
\begin{gather*}
-F''(U)qU^2\big(pU^2-2qUV +pV^2\big)+F'(U)2(qU-pV)qUV\nonumber\\
\qquad{}+
G''(V)pV^2\big(qU^2-2pUV +qV^2\big)+G'(V)2(qU-pV)pUV=0.%\label{longe}
\end{gather*}
 The equation above should hold for every value of $U$ and $V$ respectively.
The equation has the form
\begin{gather*}
-pq U^4 F''(U)
+2q^2 U^2(U  F''(U)+ F'(U)) V-pq U(U F''(U)+F'(U)) V^2\nonumber\\
\qquad{} +
\alpha(V) U^2+ \beta(V) U +\gamma(V)
=0,%\label{long}
\end{gather*}
so $F(U)$ must satisfy (necessary but not suf\/f\/icient condition) the ODE
\[
-pq U^4 F''(U)+c_2 U^2+ c_1 U + c_0=0.
\]
with some constants $c_1$, $c_2$ and $c_0$.
Similarly we get
\[
pq V^4 G''(V)
+d_2 V^2+ d_1 V + d_0
=0.\]
Checking obtained by this way solutions we obtain
\[
F(U)+G(V) =a\ln (U/V)+b(pU-qV)+c\left( \frac{p}{U}-\frac{q}{V}\right)+d
\]
and we f\/ind that for the map \eqref{F3} the following equality holds
\begin{gather}
\label{UVuv}
a\ln (U/V)+b(pU-qV)+c\left( \frac{p}{U}-\frac{q}{V}\right)=-\left[a\ln (u/v)+b(pu-qv)+c\left( \frac{p}{u}-\frac{q}{v}\right)\right].
\end{gather}

\subsection{Potentials of the Idea systems. Idolons}

Returning to equations on the lattice (by means of \eqref{id}) one can rewrite \eqref{UVuv} as
\begin{gather*}
%\label{UVuvl}
(T_2+1) \left(a \ln u+bpu+ c  \frac{p}{u}+d\right)=(T_1+1)\left(a \ln v+bqv+ c  \frac{q}{v}+d\right).
\end{gather*}
It means  there exists function $\psi$
such that
\begin{gather}
\label{pd}
a \ln u+ p \left(b u+c \frac{1}{u}\right)+d=\psi_1+\psi, \qquad
a \ln v+ q \left(b v+c \frac{1}{v}\right)+d=\psi_2+\psi
\end{gather}
where $a$, $b$, $c$ and $d$ are arbitrary constants (we assume that one of the constants $a$, $b$, $c$ is not equal zero).
The constant $d$ can be always removed by redef\/inition $\psi \rightarrow \psi +\frac{1}{2}d$ and we neglect it
\begin{gather}
\label{pdwd}
a \ln u+ p \left(b u+c \frac{1}{u}\right)=\psi_1+\psi, \qquad
a \ln v+ q \left(b v+c \frac{1}{v}\right)=\psi_2+\psi.
\end{gather}
System \eqref{pd}  and Idea system \eqref{uv} give rise to
\begin{gather}
\label{p}
\psi_{12}= \psi+a \ln \frac{pu-qv}{qu-pv} +\big(p^2-q^2\big) \left[ b \frac{uv}{qu-pv} -c \frac{1}{pu-qv} \right],
\end{gather}
so we get three parameter family of equations.
Note that in general, \eqref{uv} does not follows from~\eqref{pd} and~\eqref{p} and therefore  we will treat \eqref{uv} as an additional condition that must be satisf\/ied.
As we have said in the introduction, choice of parameters $b=0=c$ leads to equation $H3^0$~\eqref{H3}
whereas choice of parameters either $a=0=b$ or $a=0=c$ leads to equation $A1^0$~\eqref{A1}. Every such potential representation of the Idea system we refer to as idolon of the Idea system.
To the end let us write another idolon.
Namely, $a=0$ yields the equation
\begin{gather*}
\frac{\psi_{2}-\psi_{1}}{\psi_{12}-\psi}=\frac{p^2+q^2}{p^2-q^2}-\frac{pq}{p^2-q^2}\left(\frac{u}{v}+\frac{v}{u}\right),
\end{gather*}
where $u$ and $v$ are solutions of the following quadratic equations
\begin{gather*}
p \big(b u^2+c \big)=(\psi_1+\psi)u, \qquad
q \big(b v^2+c \big)=(\psi_2+\psi)v
\end{gather*}
and we still assume that \eqref{uv} holds.

\subsection[Extension to multidimension, multidimensional consistency of idolons of $I_{\rm III}$]{Extension to multidimension, multidimensional consistency\\ of idolons of $\boldsymbol{I_{\rm III}}$}

The system \eqref{uv} can be extended to multidimension. We denote by $s^i$ (mind superscript!)  function given on edges in $i$-th direction of the $\mathbb{Z}^n$ lattice,
by subscript we denote forward shift in indicated direction. The extension reads
\begin{gather}
\label{I3}
(I_{\rm III}): \quad    s^i_j = s^j \frac{p^i s^i - p^j s^j}{p^j s^i - p^i s^j}, \qquad  i,j=1, \ldots , n ,\qquad i \neq j,
\end{gather}
where $p^i$ is given function and can depend only on $i$-th independent variable.

The crucial fact is the system is compatible
\begin{gather}
\label{compat}
s^i_{jk}=s^i_{kj}.
\end{gather}
Moreover, we have
\begin{gather}
\label{ident}
(T_j+1)\left[a \ln s^i+ p^i \left(b s^i+c \frac{1}{s^i}\right)\right]=(T_i+1)\left[a \ln s^j+ p^i \left(b s^j+c \frac{1}{s^i}\right)\right].
\end{gather}
It means that there exists scalar function $\psi$ such that
\begin{gather}
\label{df}
a \ln s^i+ p^i \left(b s^i+c \frac{1}{s^i}\right)=\psi_i+\psi, \qquad  i=1, \ldots , n.
\end{gather}
From \eqref{I3} and \eqref{df} we infer that
\begin{gather}
\label{pm}
\psi_{ij}=  \psi+a \ln \frac{p^is^i-p^js^j}{p^js^i-p^is^j} +[(p^i)^2-(p^j)^2] \left[ b \frac{s^is^j}{p^js^i-p^is^j} -c \frac{1}{p^is^i-p^js^j} \right],\\
  i,j=1, \ldots , n ,\qquad i \neq j,
\nonumber
\end{gather}
where $s^i$ and $s^j$ are given implicitly by means of \eqref{df}.
Due to \eqref{compat} the system~\eqref{pm} is multidimensionaly consistent (compatible)  and we clarify what we mean by that in the following theorem  (by i-th initial line we understand in what follows the set $l^i=\{(m_1,\ldots,m_n)\in {\mathbb Z}^n \, | \, \forall \, k \neq  i :\, m_k=0 \}$
and by set of initial lines we mean $l=l^1 \cup \cdots \cup l^n$)
\begin{theorem}\label{theorem1}
For  arbitrary initial condition  on initial lines $\psi(l)$
there exists solution $($we do not exclude singularities$)$ $\psi$ of the multidimensional system \eqref{df}, \eqref{pm}  that obeys \eqref{I3}.
\end{theorem}

\begin{proof}
Indeed, take arbitrary initial condition on initial lines $\psi(l)$.
Then choose a solution~$s^i(l^i)$ (in general the value of $s^i$ is  given on the edge between vertices  that $\psi$ and $\psi_i$ are given on) of the equation
\begin{gather}
\label{dfo}
a \ln s^i(l^i)+ p^i \left(b s^i(l^i)+c \frac{1}{s^i(l^i)}\right)=\psi_i(l^i)+\psi(l^i), \qquad  i=1, \ldots , n
\end{gather}
(this is a place when non-uniqueness may  enter). We treat $s^i(l)$ as initial conditions for the system~\eqref{I3}. Due to \eqref{compat} the solution $s^i$
of~\eqref{I3} with initial conditions $s^i(l)$  exists (we admit singularities that come from zeroes of $p^j s^i - p^i s^j$).
Now, due to identity \eqref{ident} there exists function $\psi$ such that~\eqref{df} holds and the value of $\psi$ at the intersection of initial lines is equal to initial condition at the intersection of initial lines. Since $s^i$ obeys \eqref{I3} $\psi$ satisf\/ies~\eqref{pm} as well.
Finally  $\psi$ satisf\/ies the assumed arbitrary initial condition
since  formulas~\eqref{df} at initial lines coincides with~\eqref{dfo}.
\end{proof}

We refer to the system \eqref{I3} as to n-dimensional Idea system III and that is why we have denoted it by $I_{\rm III}$.

\section{Maps}
\label{maps}

As we have already mentioned our inspiration was a survey on Yang--Baxter maps.
Our goal now is to relate our f\/indings to some results of the papers \cite{ABSf,pap2}
and justify why it makes sense to talk about the Idea systems
\begin{gather}
\label{iiii}
 s^i_j = s^j \frac{p^i s^i - p^j s^j}{p^j s^i - p^i s^j}, \qquad  i=1, \ldots , n
\end{gather}
associated with maps of type  $\rm III$ rather than single Idea system. The Idea systems are related by point transformation.

Indeed, f\/irst we perform a cosmetic point transformation
$ s^i=p^i v^i$, ${p^i}^2 \rightarrow p^i$ and we get
\begin{gather*}
 v^i_j=\frac{v^j}{p^i} \frac{p^iv^i-p^jv^j}{v^i-v^j},
\end{gather*}
which in two-dimensional case after identif\/ication analogous to the one showed on the Fig.~\ref{fig:c2-c2}
yields $F_{\rm III}$ map  of paper~\cite{ABSf}
\begin{gather}
\label{fiii}
(F_{\rm III}):    \quad     U=\frac{v}{p} \frac{pu-qv}{u-v} ,  \qquad  V=\frac{u}{q} \frac{pu-qv}{u-v}.
\end{gather}
In fact by $F_{\rm III}$ we understand equivalence class of Yang--Baxter maps (cf.~\cite{pap2})
the equations~\eqref{iiii} and~\eqref{fiii} belongs to.

Now after the point transformation
$v^i = u^i (-1)^{m_1+ \cdots +m_n}$
we get
\begin{gather*}
u^i_j=-\frac{u^j}{p^i} \frac{p^iu^i-p^ju^j}{u^i-u^j}
\end{gather*}
associated 2D map of which is
\begin{gather}
\label{cha}
(cH_{\rm III}^A):    \quad   U=-\frac{v}{p} \frac{pu-qv}{u-v} ,  \qquad  V=-\frac{u}{q} \frac{pu-qv}{u-v}.
\end{gather}
After another point transformation
$u^i = {w^i}^{(-1)^{m_1+ \cdots +m_n}} {p^i}^{\frac{1}{2}[(-1)^{m_1+ \cdots +m_n}-1]}$
we obtain
\begin{gather*}
 w^i_j=-\frac{1}{w^j} \frac{w^i-w^j}{p^i  w^i- p^j w^j}
\end{gather*}
and its associated map
\begin{gather}
\label{chb}
(cH_{\rm III}^B):     \quad   U=-\frac{1}{v} \frac{u-v}{pu-qv} ,  \qquad  V=-\frac{1}{u} \frac{u-v}{pu-qv}.
\end{gather}
Maps \eqref{cha} and \eqref{chb} are not Yang--Baxter maps but they are companions (if $f:(u,v)\mapsto (U,V)$ is involutive map then the map $(u,V)\mapsto (U,v)$ we refer to as companion of map $f$, cf.~\cite{ABSf}) of Yang--Baxter maps $H_{\rm III}^A$, $H_{\rm III}^B$ of paper~\cite{PSTV}.
The maps $H_{\rm III}^A$, $H_{\rm III}^B$ can be obtained  in two-dimensional case
by the point transformation
$u^1=x$, $u^2=-y$ and $w^1=x$, $w^2=-\frac{1}{qy}$ respectively
\begin{gather*}
x_2 =\frac{y}{p} \frac{px+qy}{x+y}, \qquad y_1 =\frac{x}{q}\frac{px+qy}{x+y}  \qquad\mbox{and}\qquad
x_2 =y\frac{qxy+1}{pxy+1}, \qquad y_1 =x\frac{pxy+1}{qxy+1}
\end{gather*}
and then by mentioned identif\/ication (see Fig.~\ref{fig:c2-c2})
\begin{gather*}
(H_{\rm III}^A):      \quad U=\frac{v}{p} \frac{pu-qv}{u-v} ,  \qquad  V=\frac{u}{q} \frac{pu-qv}{u-v},
\\
(H_{\rm III}^B):     \quad U=v\frac{quv+1}{puv+1} ,  \qquad  V=u\frac{quv+1}{puv+1}.
\end{gather*}
Idea systems $(H_{\rm III}^A)$ and $(H_{\rm III}^B)$  cannot be extended to multidimension (in the sense of~\cite{ABS}).

Finally, we list in the Table~\ref{tab} basic  identities of the maps that leads to existence of potentials of the Idea systems to illustrate
how the basis changes when one changes a map.

\begin{table}[h!]\small
\centering
\caption{Basic identities of the maps that leads to existence of potentials of the Idea system.}\label{tab}
\vspace{1mm}

\begin{tabular}{lll}
\hline
Type of the map & Example of the map   & Identities \tsep{1pt}\bsep{1pt} \\
\hline
          & $\displaystyle U=\frac{v}{p}\frac{pu-qv}{u-v}$ & $\displaystyle \frac{U}{V}=\frac{qv}{pu}$ \tsep{9pt}\\
$F_{\rm III}$ &                                  & $pU-qV=-(pu-qv)$  \\
          &  $\displaystyle V=\frac{u}{q}\frac{pu-qv}{u-v}$ & $\displaystyle \frac{1}{U}-\frac{1}{V}=-\left(\frac{1}{u}-\frac{1}{v}\right)$ \bsep{7pt}\\
\hline
 &$\displaystyle U=-\frac{v}{p}\frac{pu-qv}{u-v}$ & $\displaystyle\frac{U}{V}  =  \frac{qv}{pu}$ \tsep{9pt}\\
$cH_{\rm III}^A$ & & $pU-qV=pu-qv$ \\
 &$\displaystyle V=-\frac{u}{q}\frac{pu-qv}{u-v}$ & $\displaystyle \frac{1}{U}-\frac{1}{V}=\frac{1}{u}-\frac{1}{v}$  \bsep{7pt} \\
\hline
           & $\displaystyle U=\frac{1}{v}\frac{u-v}{qv-pu}$ & $\displaystyle \frac{U}{V}=\frac{u}{v}$ \tsep{9pt} \\
$cH_{\rm III}^B$ & & $\displaystyle pU + \frac{1}{U} - qV - \frac{1}{V} = pu+\frac{1}{u}-qv-\frac{1}{v}$ \\
          & $\displaystyle V=\frac{1}{u}\frac{u-v}{qv-pu}$ & $\displaystyle pU - \frac{1}{U} - qV + \frac{1}{V} =-
          \left(pu-\frac{1}{u}-qv+\frac{1}{v}\right)$ \bsep{7pt}  \\
\hline
& $\displaystyle U=\frac{v(pu+qv)}{p(u+v)}$& $\displaystyle \frac{U}{V}=\frac{qv}{pu}$\tsep{9pt}    \\
$H_{\rm III}^A$& &$pU+qV=pu+qv$   \\
&$\displaystyle V=\frac{u(pu+qv)}{q(u+v)}$ & $\displaystyle \frac{1}{U}+\frac{1}{V}=\frac{1}{u}+\frac{1}{v}$  \bsep{7pt} \\
\hline
& $\displaystyle {U} = v\frac{quv+1}{puv+1}$ & $UV=uv$  \tsep{9pt} \\
$H_{\rm III}^B$& & $\displaystyle pU+qV+\frac{1}{U}+\frac{1}{V}=pu+qv+\frac{1}{u}+\frac{1}{v}$  \\
&$\displaystyle {V} = u \frac{puv+1}{quv+1}$ & $\displaystyle pU-qV-\frac{1}{U}+\frac{1}{V}=-\left(pu-qv-\frac{1}{u}+\frac{1}{v}\right)$ \bsep{7pt} \\
\hline
\end{tabular}
\end{table}

\section{Hirota's KdV lattice equation}
\label{hero}

As the second example we consider
Hirota's KdV lattice equation \cite{hirota-0}
\[
x_{12}-x=\kappa \left(\frac{1}{x_2}-\frac{1}{x_1}\right).
\]
By the substitution $u=x_1x$,  $v=x_2x$,
we get
\begin{gather}
\label{kdvs}
u_2=v+\kappa \left(1-\frac{v}{u}\right), \qquad v_1=u+\kappa \left(-1+\frac{u}{v}\right).
\end{gather}
On applying identif\/ication \eqref{id}
\begin{gather}
\label{idk}
u=u(m,n),\qquad v=v(m,n), \qquad U=u(m,n+1), \qquad V=v(m+1,n)
\end{gather}
we obtain an involutive mapping associated to system \eqref{kdvs}
\begin{gather}
\label{ybhirota}
U=v+\kappa \left(1-\frac{v}{u}\right), \qquad V=u+\kappa \left(-1+\frac{u}{v}\right).
\end{gather}
Mapping (\ref{ybhirota}) satisf\/ies (this is the outcome of searching for such functions $F$ and $G$ that $F(U)+G(V)=f(u)+g(v)$ as described in the previous section):
\begin{gather*}
%\label{kdvp}
 \frac{U}{V}=  \frac{v}{u},\qquad
(U-\kappa)(V+\kappa)=(u-\kappa)(v+\kappa), \qquad
 \frac{V(U-\kappa)}{U(V+\kappa)}=\frac{v(u-\kappa)}{u(v+\kappa)},
\end{gather*}
hence (coming back to lattice variables \eqref{idk}) we can introduce  the potentials $x$, $y$ and $z$
\begin{alignat}{3}
& u=x_1 x, \qquad & & v=x_2 x,& \nonumber\\
& u-\kappa=y_1/y,\qquad && v+\kappa=y/y_2,& \nonumber\\
& \frac{u-\kappa}{u} =z_1/z,\qquad & &  \frac{v+\kappa}{v} =z_2/z.& \label{potentials}
\end{alignat}
Eliminating $u$ and $v$ from  \eqref{kdvs} we arrive at the following lattice equations
\begin{gather}
x_{12}-x=\kappa(1/x_2-1/x_1),\qquad
y_1y-y_{12}y_1=\kappa(y_{12}y+y_1y_2),   \nonumber    \\
z_{12}z+z_1z_2=z_{12}z_2+z_{12}z_1.\label{kdvi}
\end{gather}
One can treat the equations as representatives of a three-parameter family of equations on~$\phi$
\begin{gather}
\frac{\phi_{12}\phi}{\phi_{1}\phi_{2}}= \big[(u-\kappa)(v+\kappa) +\kappa^2\big]^{a (-1)^{m+n+1}-b} u^{b-c} v^{b+c},  \nonumber\\
\frac{\phi_{1}}{\phi}= u^{a (-1)^{m+n}-b} (u-\kappa)^{b+c},      \qquad
\frac{\phi_{2}}{\phi}= v^{a (-1)^{m+n}-b} (v+\kappa)^{b-c},\label{HG}
\end{gather}
corresponding to the choice of parameters $b=0=c$, $a=0=b$ and $a=0=c$ respectively.

What more important is that from \eqref{potentials} we infer
\begin{gather*}
\frac{z_1}{z}=\frac{y_1}{ x_1 x y},  \qquad  \frac{z_2}{z}= \frac{y}{y_2 x_2 x}.
\end{gather*}
Compatibility condition that guarantees existence of function $z$ reads
\begin{gather*}
\left(\frac{x_2}{x_1}\right)^2= \left(\frac{y_{12} y}{y_1 y_2}\right)^2,
\end{gather*}
from where  we get
\begin{gather*}
x= \frac{\tau_{12} \tau}{\tau_{1}\tau_{2}}, \qquad y= \frac{\tau_2}{\tau_1}, \qquad z= \frac{\tau}{ \tau_{12}}.
\end{gather*}
Eliminating $x$, $y$ and $z$ from  \eqref{potentials} we arrive at a compatible pair of bilinear forms of Hirota's KdV (cf.~\cite{NijOht})
\begin{gather*}
\tau_{112}\tau -\kappa \tau_{11}\tau_{2}=\tau_{12}\tau_{1},\qquad
\tau_{122}\tau +\kappa \tau_{22}\tau_{1}=\tau_{12}\tau_{2}.
\end{gather*}

\section{B\"acklund transformations between idolons}
\label{bt}

In both presented examples one can f\/ind B\"acklund transformation between idolons.
For instance eliminating~$u$ and~$v$  from f\/irst two lines of~\eqref{potentials} one gets B\"acklund transformation between f\/irst two equations
of~\eqref{kdvi}
\[
\frac{y_1}{y}=x_1x-k, \qquad \frac{y}{y_2}=x_2x+k.
\]
Similarly in the case of $I_{\rm III}$ one can obtain B\"acklund transformation \eqref{xy}.

 Finally, we  present the  B\"acklund transformation between $A1^0$ (\ref{A1}) and the ido\-lon~\eqref{pdwd},~(\ref{p}). Namely, if $y$ satisf\/ies $A1^0$
then
\begin{itemize}\itemsep=0pt
\item
function $\psi$ given by
\begin{gather*}
a \ln{\frac{p}{y_1+y}}+\frac{b p^2}{y_1+y}+c (y_1+y)=\psi_1+\psi, \\
a \ln{\frac{q}{y_2+y}}+\frac{b q^2}{y_2+y}+c (y_2+y)=\psi_2+\psi
\end{gather*}
exists (compatibility conditions are satisf\/ied due to the fact that $y$ satisf\/ies $A1^0$);
\item
functions $u$ and $v$ given by $u=\frac{p}{y_1+y}$, $v=\frac{q}{y_2+y}$ obey Idea system \eqref{uv};
\item
function $\psi$ obeys \eqref{pdwd}, \eqref{p}.
\end{itemize}

 \section{Conclusions}
\label{ccl}

 In this paper we focused on two 3-parameter families of lattice equations. The f\/irst one~(\ref{ip}),~\eqref{ipd} and \eqref{u2v1} is related to mappings of type III which were introduced in \cite{ABSf,pap2}. Two members (idolons) of the later are, the Hirota's sine-Gordon equation  and the
lattice Schwarzian KdV~\cite{NC}  in a disguise. Generally, all idolons are connected through B\"acklund transformations and they are 3D-consistent in the sense described in the paper.
 In the not-too-distant future we are \mbox{going} to investigate families of equations related to given integrable systems not only by discrete quadratures but also by B\"acklund transformation from the Def\/inition~\ref{BTdef}.

The second family described by (\ref{HG}) and \eqref{kdvs} is not  3D-consistent. Nevertheless, all of its idolons are connected through B\"acklund transformations, and since an idolon of this family is the Hirota's KdV equation, the whole family   inherits some integrability properties e.g.\
$\tau$-function formulation.

We would like to emphasize once more that the main object under consideration are Idea systems \eqref{uv} (or its $n$-dimensional version \eqref{I3}) and \eqref{kdvs}. The main observation is that the Idea systems  admit three-dimensional vector space of scalar potentials (formulas \eqref{pdwd} in  case of two-dimensional Idea $I_{\rm III}$ and \eqref{df} in  the n-dimensional case, see also second and third formulas of \eqref{HG}). In a forthcoming paper we will discuss all Idea systems that arise from equations of Adler--Bobenko--Suris list.
   In other words, we plan to investigate all mappings in~\cite{ABSf,pap2}, determine their associate Idea systems and put more light into  integrability properties of the associated family of lattice equations. Also, it will be interesting to investigate the mappings that arise when one imposes periodic staircase initial data on these families of lattice equations.  Another objective  is to derive the  discrete Painlev{\'e} equations associated with these families.

Finally, we will discuss  the case of real-valued functions, which can lead to  standard 3D-consistent lattice equations.  For instance  for the  idolon  we proposed  in \cite{KaNie}
 \begin{gather}
 \label{lcub1}
 f_{12}=f+{  (p-q)\left[v-u+\frac{f_1-f_2}{(u-v)^2} +\frac{(p-q)^2}{(u-v)^3}\right]},\\
  %\label{lcub2}
u^3 +a u =  f_1-f , \qquad   v^3+bv = f_2-f, \qquad a-b=3(q-p).\nonumber
\end{gather}
 assuming
 $f:   \mathbb{Z}^2 \rightarrow \mathbb{R}$ and $a,b>0$ the only  real solutions of the cubic equations are
\begin{gather}
u={\sqrt[3]{\frac{f_1-f}{2}+\sqrt{ \frac{(f_1-f)^2}{4}+\frac{a^3}{27}} }+\sqrt[3]{\frac{f_1-f}{2}-\sqrt{ \frac{(f_1-f)^2}{4}+\frac{a^3}{27}} }},
\nonumber\\
v={ \sqrt[3]{\frac{f_2-f}{2}+\sqrt{ \frac{(f_2-f)^2}{4}+\frac{b^3}{27}} }+\sqrt[3]{\frac{f_2-f}{2}-\sqrt{ \frac{(f_2-f)^2}{4}+\frac{b^3}{27}} }}.\label{realcub}
\end{gather}
Then the real lattice equation (\ref{lcub1}), with $u$ and $v$ given by (\ref{realcub}), is 3D-consistent.

From another perspective, instead of dealing with the family of lattice equations, it seems more fundamental to def\/ine a model that consists of the Idea system and the associate potential equation (e.g.\ equations~(\ref{uv}), (\ref{pd}) or (\ref{I3}), (\ref{df}) for the multidimensional extension). Then the family of 3D-consistent (see Theorem~\ref{theorem1}) lattice equations follows naturally. But what more important, this is a new lattice model, def\/ined in both vertices and edges of a 2D square lattice ($n$D lattice in the multidimensional extension).  Such models have also been  introduced in the recent work of Hietarinta and Viallet~\cite{hi-via}.

\subsection*{Acknowledgements}
 We would like to thank  organizers of SIDE-9 conference in Varna for their hospitality and f\/inancial support.
Special thanks to Georgi Grahovski for showing us the other side of Varna. M.N.\ thanks Frank Nijhof\/f for pointing papers~\cite{Ni1,Ni2}.

\pdfbookmark[1]{References}{ref}
\LastPageEnding

\end{document}